 \documentclass[acmsmall]{acmart}

 \setcopyright{none}

\usepackage{multirow}
\usepackage{graphicx}
\begin{document}

\title{Exploring the Impact of Emotional Voice Integration in Sign-to-Speech Translators for Deaf-to-Hearing Communication}

%
\author{Hyunchul Lim}
\email{hl2365@cornell.edu}
\affiliation{%
  \institution{Cornell University}
  \city{Ithaca}
  \state{New York}
  \country{USA}
  \postcode{14850}
}

\author{Minghan Gao}
\email{mg2328@cornell.edu}
\affiliation{%
  \institution{Cornell University}
  \city{Ithaca}
  \state{New York}
  \country{USA}
  \postcode{14850}
}

\author{Franklin Mingzhe Li}
\email{mingzhe2@cs.cmu.edu}
\affiliation{%
  \institution{Carnegie Mellon University}
  \city{Pittsburgh}
  \state{Pennsylvania}
  \country{USA}
  \postcode{14850}
}

\author{Nam Anh Dang}
\email{nd433@cornell.edu}
\affiliation{%
  \institution{Cornell University}
  \city{Ithaca}
  \state{New York}
  \country{USA}
  \postcode{14850}
}

\author{Ianip Sit}
\email{ss4521@rit.edu}
\affiliation{%
  \institution{Rochester Institute of Technology}
  \city{Rochester}
  \state{New York}
  \country{USA}
  \postcode{14850}
}

\author{Michelle M Olson}
\email{mmo9420@g.rit.edu}
\affiliation{%
  \institution{Rochester Institute of Technology}
  \city{Rochester}
  \state{New York}
  \country{USA}
  \postcode{14850}
}
\author{Cheng Zhang}
\email{chengzhang@cornell.edu}
\affiliation{%
  \institution{Cornell University}
  \city{Ithaca}
  \state{New York}
  \country{USA}
  \postcode{14850}
}

%
\renewcommand{\shortauthors}{Lim, et al.}

\begin{abstract}
Emotional voice communication plays a crucial role in effective daily interactions. Deaf and hard-of-hearing (DHH) individuals often rely on facial expressions to supplement sign language to convey emotions, as the use of voice is limited. However, in American Sign Language (ASL), these facial expressions serve not only emotional purposes but also as linguistic markers, altering sign meanings and often confusing non-signers when interpreting a signer’s emotional state. Most existing ASL translation technologies focus solely on  signs, neglecting the role of emotional facial expressions in the translated output (e.g., text, voice). This paper present studies which 1) confirmed the challenges for non-signers of interpreting emotions from facial expressions in ASL communication,  of facial expressions, and 2) how integrating emotional voices into translation systems can enhance hearing individuals’ comprehension of a signer’s emotions. An online survey conducted with 45 hearing participants (Non-ASL Signers) revealed that they frequently misinterpret signers’ emotions when emotional and linguistic facial expressions are used simultaneously. The findings indicate that incorporating emotional voice into translation systems significantly improves the recognition of signers' emotions by 32\%. Additionally, further research involving 6 DHH participants discusses design considerations for the emotional voice feature from both perspectives, emphasizing the importance of integrating emotional voices in translation systems to bridge communication gaps between DHH and hearing communities.
\end{abstract}

\maketitle

\section{Introduction}

The Deaf and Hard-of-Hearing (DHH) community relies on various sign languages, such as American Sign Language (ASL), for daily communication. To bridge communication gaps between DHH individuals and non-signing hearing people, numerous sign language translation systems have been developed to translate signed language into text or speech. These systems enable hearing individuals to understand sign language, fostering greater accessibility and inclusivity in communication contexts \cite{grover2021sign, dreuw2010signspeak, fang2017deepasl}. However, existing translation systems primarily focus on conveying the literal meaning of signs, overlooking the emotional nuances that are integral to effective communication.

A key limitation of current sign-to-speech/text translation systems \cite{babour2023intelligent, san2008speech, farooq2021advances, bragg2019sign, fang2017deepasl} is their inability to convey the signer’s emotional information. Typically, these systems employ a neutral, machine-generated voice for audio output, resulting in translations that fail to capture the emotional undertones of the signer's expressions. This absence of emotional cues can impair hearing individuals' ability to interpret the signer’s intended emotions accurately, leading to potential miscommunication. Non-signing hearing individuals often rely on multimodal signals, including facial expressions and contextual cues, to interpret emotions \cite{wehrle2000studying}. Yet, comprehending emotions from a single-sentence-based translation system without emotional nuance demands substantial cognitive effort, which may hinder effective understanding.

Interpreting emotions in sign language is challenging for non-signing hearing individuals for several reasons. First, emotional expressions are complex and vary significantly across individuals \cite{elliott2013facial}. Second, facial expressions in sign language often serve dual purposes—conveying emotions and fulfilling linguistic functions, such as marking emphasis or intensity \cite{adolphs2002recognizing, shaffer2018exploring, huenerfauth2011evaluating}. For instance, a linguistic marker like 'INTENSE' may be misinterpreted as anger, even when the signer’s actual emotional state is neutral. Moreover, when both emotional and linguistic facial expressions are used concurrently (e.g., "It's snowing a ton in New York" signed as 'HERE NEW-YORK [Happy] SNOW [INTENSE]'), hearing individuals may struggle to distinguish between these signals. Compounding this challenge is the mismatch between neutral voice output and emotional cues in facial expressions, which can create cognitive dissonance and further obscure emotional understanding \cite{nass2001does}.

To address these unexplored areas, this paper investigates the following research questions using three different translation systems providing: (A) subtitles only, with no voice; (B) a neutral voice; and (C) emotional voices.
\begin{itemize}
  \item (RQ1) How do non-ASL hearing individuals recognize signers’ emotions when observing both emotional facial expressions and translation information in a communication scenario?;
  \item (RQ2) How do non-ASL hearing individuals recognize signers’ emotions when observing linguistic
facial expressions of a signer?;
  \item (RQ3) How do non-ASL hearing individuals recognize signers’ emotions when observing multiple facial
expressions such as both emotional and linguistic facial expression simultaneously during signing?;
\item (RQ4) How should emotional voice be designed for emotional-voice translation systems to aid communication between hearing and DHH?
\end{itemize}

To explore RQ1-3, we conducted three online surveys with 45 participants. In each survey, participants watched video clips of a signer, identified the signer’s emotions and the meaning of signs, and rated their mental effort in selecting their answers. Each survey provided the same content but differed in the type of translation information: 1) translated sign language text with no voice (only subtitles), 2) a neutral voice, and 3) an emotional voice. We found that non-ASL hearing participants had greater difficulty recognizing emotions when interpreting both facial expressions and translation. Additionally, linguistic facial expressions negatively impacted their perception of the signer’s emotions, increasing mental effort. However, the resutls showed providing an emotional voice improved recognition performance, even when linguistic facial expressions contradicted the signer’s emotions. 

To address RQ4, we collected open-ended feedback on the design of emotional voices from 6 DHH users and 45 hearing participants for future emotional voice ASL translation systems. This feedback revealed ranked emotional voice types by importance, suggested additional emotional voice options, and desired features for emotional voice. Overall, we believe our findings provide an in-depth understanding of how different types of translation systems affect hearing individuals' ability to recognize signers' emotions in real-world communication scenarios, emphasizing the importance of emotional voice in translation systems. Our results offer insights for HCI and CSCW researchers into the design of future ASL translation systems (e.g., incorporating emotional voice) by highlighting practices and challenges associated with various translation types, thereby enhancing effective communication between hearing and DHH individuals

Our work builds upon and bridges previous CSCW research focused on multimodal communication, emotional expression in collaborative contexts, and accessibility. While much of this research has examined emotional expression in digital communication to enhance mutual understanding and reduce cognitive dissonance, few studies address the unique challenges faced by DHH individuals in cross-modal communication. We specifically tackle the critical gap concerning the lack of emotional nuance in ASL translation systems used by hearing, non-signing individuals. By integrating emotional voice for DHH-hearing interactions, our work offers a novel perspective on how CSCW principles of emotional and communicative alignment can guide the design of accessible, emotionally aware systems



\section{Background and Related Work}

\subsection{Facial Expression in American Sign Language}
Facial expressions play a central role in ASL, fulfilling both emotional and grammatical functions \cite{shaffer2018exploring, mccullough2005neural, corina1999neuropsychological}. Unlike spoken languages, where tone conveys nuance, ASL relies on facial expressions to enrich and clarify signed content. ASL signers use expressions to convey emotions and linguistic information, such as marking questions \cite{mccullough2009categorical}, indicating intensity \cite{freitas2017grammatical, corina1999neuropsychological}, or altering meanings \cite{reilly1990faces}. For example, eyebrow movements can signal questions, while puffed cheeks indicate intensity. These markers are essential to ASL grammar and generally lack emotional intent, which can lead hearing non-signers to misinterpret them as emotional cues \cite{mccullough2009categorical, corina1999neuropsychological, freitas2017grammatical}. Studies on linguistic facial expressions have primarily focused on native signers and DHH individuals \cite{mccullough2005neural, reilly2006faces, elliott2013facial} or recognition technology \cite{michael2010computer, shaffer2018exploring}, with limited research on non-signer hearing individuals' perspectives. Although one study highlights hearing non-signers’ difficulty classifying emotional and linguistic expressions \cite{mccullough2009categorical}, no research, to our knowledge, examines whether hearing non-signers interpret linguistic expressions as emotions or how they perceive simultaneous emotional and linguistic expressions in communicative contexts. This paper investigates linguistic marker perceptions as emotions, related challenges, and assesses an AI-generated emotional voice translation system to address these issues.

\subsection{Emotion Perception from Face, Voice, and Contextual Information}
Emotional facial expressions are crucial to nonverbal communication, fostering collaborative interactions \cite{hess2016nonverbal}. For hearing individuals, emotional expressions are multimodal, combining visual, auditory, and spatial-temporal contexts. Initial facial expression research relied on static images \cite{creed2008psychological}; however, dynamic \cite{rodger2021recognition} and multimodal stimuli \cite{de2000perception} are now commonly explored. While dynamic stimuli's effectiveness remains debated, with some studies showing no significant difference compared to static images \cite{gold2013efficiency, fiorentini2011there}, hearing individuals generally perceive emotions through both facial expressions \cite{ekman1971universals} and speech \cite{murray1993toward}. De Gelder et al. \cite{de2000perception} showed participants identified emotions faster in combined face-and-voice settings than in face-only or audio-only settings, suggesting a multimodal approach to emotional perception.

Emotion-voice mismatches occur when audio tones do not align with a signer’s visual expressions, negatively impacting emotion perception \cite{creed2008psychological}. This misalignment, particularly evident in ASL when linguistic expressions are mistaken for emotions, can confuse non-signers and complicate interpretation. Our study investigates this issue and evaluates whether emotional voice support can help resolve these mismatches. Furthermore, real-world emotion perception depends heavily on situational context, which remains underrepresented in many studies \cite{lee2012context, carroll1996facial, wallbott1988and}. Emotion recognition in ASL settings, particularly regarding signers' expressions in realistic contexts, remains largely unexamined. Effective communication between DHH and hearing individuals requires both accurate interpretation of signs and understanding of the signer’s emotions. This study explores how hearing participants perceive signers’ emotions with and without translation to improve emotion perception in real-world communication.

\subsection{Sign Language Translation Systems}
Sign language translation systems aim to facilitate communication between DHH individuals and hearing individuals by converting signed language into speech or text. These systems leverage various technologies, including motion capture \cite{havasi2005motion}, machine learning \cite{jin2023smartasl, jin2023transasl}, and natural language processing \cite{adigwe2018emotionalvoicesdatabasecontrolling}, to accurately recognize and translate signs. Recent advancements have also introduced facial expression recognition using \cite{li2024eyeecho, li2013simultaneous, chen2021neckface, soleymani2015analysis, soleymani2014continuous, cohen2003facial}, enhancing the capture of non-manual markers such as mouthing, which are essential for conveying grammatical and semantic information \cite{jin2023smartasl, jin2023transasl}.

However, many current technologies focus primarily on recognizing manual markers, often overlooking these crucial non-manual aspects. For example, effective ASL translation systems can integrate detailed facial expression tracking, improving the accuracy of translating linguistic nuances in ASL, including negative head shaking and question markers \cite{jin2023smartasl, jin2023transasl}. Recent research \cite{song2018gesture} has explored the incorporation of emotional facial expressions into translation systems, assessing the effectiveness of synthesized emotional speech. AI-generated emotional voice technology \footnote{https://azure.microsoft.com/en-us/blog/announcing-new-voices-and-emotions-to-azure-neural-text-to-speech/} \cite{liu2021reinforcement, im2022emoq} has also been extensively developed to produce varied vocal expressions \cite{adigwe2018emotionalvoicesdatabasecontrolling, roshan2024sentient}. While this technology is advancing towards implementation in ASL translation systems, there remains a significant gap in understanding how to design emotional voice into these systems effectively, especially in communication scenarios involving DHH and hearing individuals. In this paper, we evaluate current translation systems in simulated communication settings and explore what is needed to address existing challenges in designing future ASL translation systems.

\section{
Method}


\subsection{Survey Outline}
\begin{figure}[h]
  \includegraphics[width=\linewidth]{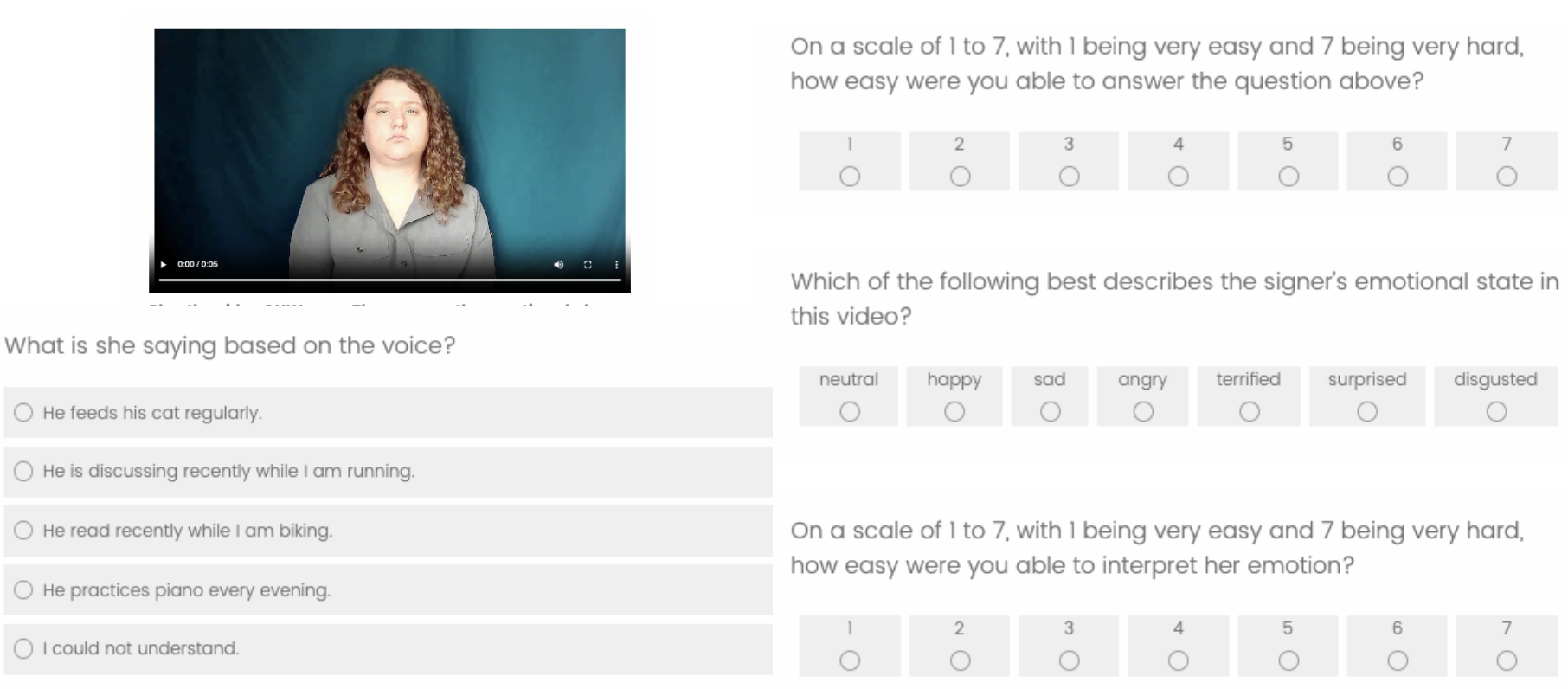}
  \caption{Example Survey Questions: After watching a video clip once, participants were asked to answer the questions related to .}
  \Description{}
  \label{fig:survey}
\end{figure}

The survey for non-ASL hearing participants began with demographic questions, including gender and age. They were asked to watch video clips where the sign model perform emotional and/or linguistic facial expression with/without signing, and select the meaning of the signs and associated emotions, rating the difficulty of recognitions on a scale of 1 to 7 (See Fig. \ref{fig:survey}). Participants also were asked to rank seven emotions by importance, providing their reasons (\textit{i.e., "Which emotion is more critical to be interpreted. Please rank them and explain your rating"}), and suggest features for designing emotional voices in ASL translation systems (i.e., \textit{"Are the 7 emotions (neutral, happy, sad, angry, terrified, disgust, surprised) provided sufficient, or should other emotions be included?"}). 

The survey for DHH participants also began with demographic questions and included an explanation of an emotional-voice translation system. They then were asked to rank seven emotions by importance (same question above), providing their reasons, and provide feedback on essential features (same question above) for designing emotional voices to facilitate communication with non-ASL hearing individuals.





\subsection{Survey Design}

\subsubsection{Facial Expressions}
\begin{figure}[h]
  \includegraphics[width=\linewidth]{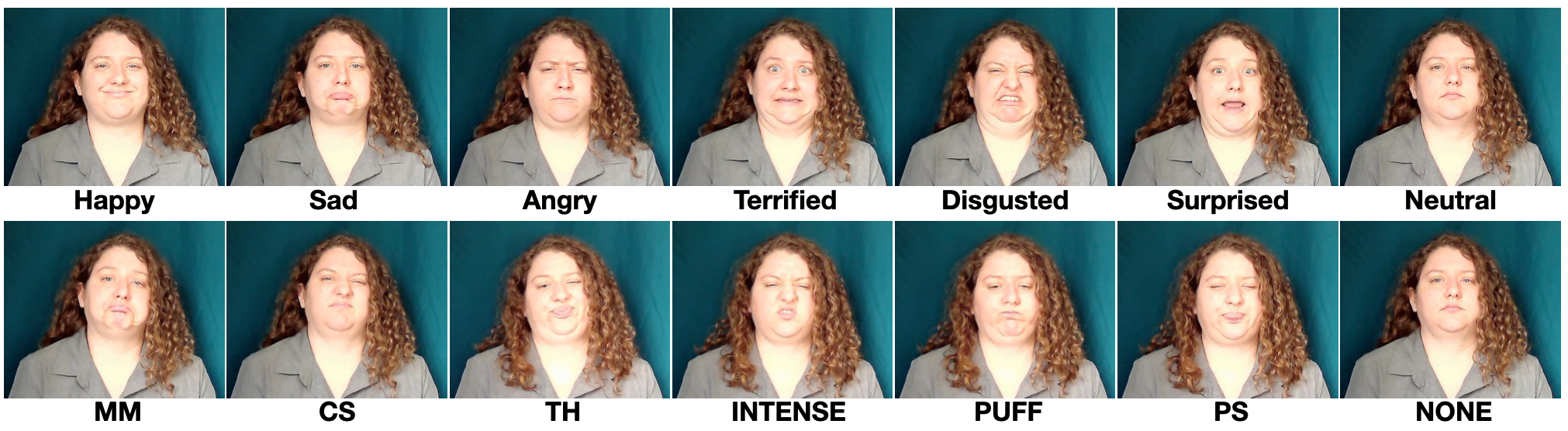}
  \caption{Facial expressions in ASL convey emotions (top) and provide additional information, such as altering the meaning of signs, serving as linguistic markers (bottom).}
  \Description{}
  \label{fig:facialexpression}
\end{figure}

In ASL, facial expression has been used to express emotions and also linguistic makers to change the meaning of the signs (See. Fig.\ref{fig:facialexpression}). Based on previous work \cite{ekman1971constants}, we selected seven basic emotions: happy, sad, angry, terrified, surprised, disgusted, plus neutral. Additionally, seven common linguistic markers were chosen (see below) based on prior research \cite{shaffer2018exploring}. The linguistic facial expression changes the meaning of signs. For example, without linguistic markers, "HE DRIVE CAR" translates to "He drove a car." However, with a linguistic facial expression, e.e., MM, the same sign conveys a different meaning, such as "He \textit{effortlessly} drove a car". Some emotional and linguistic facial expressions appear similar, though they are performed differently \cite{corina1999neuropsychological}.

\begin{itemize}
    \item MM : lips pressed together while protruded
    \begin{itemize}
        \item meaning effortlessly or regularly, e.g. He \textit{effortlessly} drove a car. 
    \end{itemize}
 
    \item CS : creating a double chin with slightly narrow eyes as if skeptical while slightly frown
     \begin{itemize}
    \item meaning recently, e.g., They \textit{recently} adopted a puppy. 
      \end{itemize}
    \item TH : tongue protrudes slightly    \begin{itemize}
        \item
        meaning carelessly or inattentively, e.g. He \textit{carelessly} spilled the soup.
        \end{itemize}
    \item INTENSE : furrowed eyebrows, grimaced mouth   \begin{itemize}
        \item
        meaning much greater than expected, e.g. It’s snowing \textit{a ton} in New York.
        \end{itemize}
    \item PUFF : puffed cheeks    \begin{itemize}
        \item
        meaning a great deal or a large amount, e.g. She \textit{really} worked out at the gym.
        \end{itemize}
    \item PS : tightened lips, slightly frowning eyebrows   \begin{itemize}
        \item
        meaning smoothly, very thin, quickly, or easily, e.g. She ran \textit{quickly}.
        \end{itemize}
    \item NONE: meaning nothing.  
    
 \end{itemize}
 





\subsubsection{ASL Sign Language Dataset}

To systematically explore how non-ASL hearing people can understand signer's emotion during signing, we created 98 sign language sentences using multiple facial expressions and different order of facial expressions with linguistic markers. 




\begin{equation}
7 \text{ linguistic markers} \times 7 \text{ emotions} \times 2 \text{ orders of facial expressions} = 98 \text{ sentences}
\end{equation}


The sentences were designed for experimental purpose to include one emotion and one linguistic marker for each sentence. To fulfill the compatibility with all emotions and linguistic markers, we simplified the sentences to only include two subjects and two verbs. For example, the sign language with facial expressions "He drive (MM), I read (Angry)" is translated into English sentence "He is driving while I am reading." with subtitles or speech. The verbs were selected as WRITE, DRIVE, READ, BICYCLE, SIGN, STUDY, RUN, CARRY, DISCUSS, and OPEN. These verbs were selected because they can be used naturally in conjunction with all of the emotional and linguistic facial expressions. \cite{mccullough2005neural}

The structure of each sentence incorporates a linguistic marker with the first verb and an emotional expression with the second, with variants provided for different sequencing. This dual structuring allows an exploration into how the sequence of emotional expression and linguistic markers can influence the perceived emotional tone of the sentences.
\begin{itemize}
    \item Order A: Linguistic marker first, followed by emotional expression. e.g., He drive (MM), I read (Happy)
    \item Order B: Emotional expression first, followed by linguistic marker.e.g, I read (Angry), He drive (CS)
\end{itemize}



\subsubsection{AI Emotion-based Voice Generator}
To generate emotional voice, we used an AI emotional voice generator, which we evaluated for quality with 19 non-ASL hearing participants before incorporating it into the survey.

We used Azure Neural Text to Speech (Azure Neural TTS) \footnote{https://azure.microsoft.com/en-us/blog/announcing-new-voices-and-emotions-to-azure-neural-text-to-speech/} for generating emotional speech. In our experiment, we used five female emotional voices such as neutral, excited (happy),  sad, angry, and terrified.  Since Azure does not provide the disgusted and surprised voice, we generated these voices using a text-to-speech (TTS) model.  To synthesize the female disgusted and surprised voices, we used the Piper Local Text to Speech system\footnote{https://github.com/rhasspy/piper}, specifically the lessac-medium model \footnote{https://huggingface.co/datasets/rhasspy/piper-checkpoints/commit/19ca249c3d7c490dbbefbaf775f74df10681d9a4} with the disgusted voice lines from the EmoV-DB Database \footnote{https://github.com/numediart/EmoV-DB}\cite{adigwe2018emotionalvoicesdatabasecontrolling} and the surprised voice lines from ESD Database \footnote{https://hltsingapore.github.io/ESD/index.html}\cite{zhou2021seen}. Specifically, we used 333 disgust voice samples ranging from 3-10 seconds from speaker "Spk-Bea" in the EmoV-DB Database and 350 surprise utterances of 2-3 seconds from speaker "0017" in the ESD Database.





We conducted an online survey with 19 participants to determine which voice was most suitable for expressing various emotions, given that Azure provides multiple voice options. Participants evaluated how well each voice expressed specific emotions and rated the effectiveness on a scale of 1 to 5, where 1 represented "bad" and 5 represented "excellent". The survey consisted of 203 questions and took approximately 20 minutes to complete, with no compensation provided. Participation was voluntary. Based on the results (See Table \ref{table:voice_evaluation}), we selected voices with an average rating above 3.5. For neutral emotion, we chose en-US-Jenny (M: 3.88, SD: 0.89); for happy emotion, en-US-Aria (M: 3.6, SD: 0.86); for sad emotion, en-US-Sara (M: 3.68, SD: 0.98); for angry emotion, en-US-Nancy (M: 3.65, SD: 0.99); and for terrified emotion, en-US-Jane (M: 4.12, SD: 0.90). For surprised and disgusted emotions, we chose Bea (M: 3.73, SD: 1.17) for disgusted and 0018 (M: 4.12, SD: 0.90) for surprised.

\begin{table}[h]
\caption{Evaluation of AI-Generated Emotional Voices: Emotional voices was rated on a scale of 1 to 5, where 1 represented "poor" and 5 represented "excellent," with the mean and standard deviation in parentheses. The bold one was selected from the same emotional voices for vidoe clips.}
\scalebox{0.85}{
\begin{tabular}{c|c|c|c|c|c|c|c|c|c}
\hline
                  & Aria                                                   & Jane                                                   & Jenny                                                  & Nancy                                                  & Sara                                                   & Jenie                                                 & Bea                                                    & "0017"                                                 & "0018"                                                \\ \hline
Neutral \cite{AzureNeuralTTS}           & \begin{tabular}[c]{@{}c@{}}3.51  \tiny(0.95)\end{tabular} & \begin{tabular}[c]{@{}c@{}}3.5  \tiny(0.92)\end{tabular}  & \begin{tabular}[c]{@{}c@{}}\textbf{3.88 } \tiny(0.89)\end{tabular} & \begin{tabular}[c]{@{}c@{}}3.6  \tiny(0.98)\end{tabular}  & \begin{tabular}[c]{@{}c@{}}3.82  \tiny(0.89)\end{tabular} & -                                                     & -                                                      & -                                                      & -                                                     \\ \hline
Happy \cite{AzureNeuralTTS}           & \begin{tabular}[c]{@{}c@{}}\textbf{3.6}  \tiny(0.86)\end{tabular}  & \begin{tabular}[c]{@{}c@{}}3.59  \tiny(0.96)\end{tabular} & \begin{tabular}[c]{@{}c@{}}3.46  \tiny(0.91)\end{tabular} & \begin{tabular}[c]{@{}c@{}}3.13 \tiny(1.03)\end{tabular} & \begin{tabular}[c]{@{}c@{}}3.52  \tiny(0.89)\end{tabular} & -                                                     & -                                                      & -                                                      & -                                                     \\ \hline
Sad \cite{AzureNeuralTTS}           & \begin{tabular}[c]{@{}c@{}}3.13  \tiny(1.08)\end{tabular} & \begin{tabular}[c]{@{}c@{}}3.24  \tiny(1.07)\end{tabular} & \begin{tabular}[c]{@{}c@{}}3.44  \tiny(1.05)\end{tabular} & \begin{tabular}[c]{@{}c@{}}3.56  \tiny(1.04)\end{tabular} & \begin{tabular}[c]{@{}c@{}}\textbf{3.68}  \tiny(0.98)\end{tabular} & -                                                     & -                                                      & -                                                      & -                                                     \\ \hline
Angry \cite{AzureNeuralTTS}          & \begin{tabular}[c]{@{}c@{}}2.97  \tiny(1.03)\small \end{tabular} & \begin{tabular}[c]{@{}c@{}}3.55 \tiny(0.93)\end{tabular} & \begin{tabular}[c]{@{}c@{}}3.38  \tiny (1.01)\small \end{tabular} & \begin{tabular}[c]{@{}c@{}}\textbf{3.65}  \tiny(0.99)\small \end{tabular} & \begin{tabular}[c]{@{}c@{}}3.04\tiny (1.12)\small \end{tabular} & -                                                     & -                                                      & -                                                      & -                                                     \\ \hline
Terrified \cite{AzureNeuralTTS}       & \begin{tabular}[c]{@{}c@{}}3.07  \tiny(1.11)\small \end{tabular} & \begin{tabular}[c]{@{}c@{}} \textbf{4.12} \tiny  (0.9)\end{tabular}  & \begin{tabular}[c]{@{}c@{}}3.67  \tiny(0.98)\small \end{tabular} & \begin{tabular}[c]{@{}c@{}}3.63 \tiny(1.12)\small \end{tabular} & \begin{tabular}[c]{@{}c@{}}3.0 \tiny(1.1)\small \end{tabular}     & -                                                     & -                                                      & -                                                      & -                                                     \\ \hline
Disgusted \cite{adigwe2018emotionalvoicesdatabasecontrolling}          & -                                                      & -                                                      & -                                                      & -                                                      & -                                                      & \begin{tabular}[c]{@{}c@{}}3.28 \tiny(1.19)\small \end{tabular} & \begin{tabular}[c]{@{}c@{}}\textbf{3.73} \tiny(1.17)\small \end{tabular} & -                                                      & -                                                     \\ \hline
Surprised  \cite{zhou2021seen}          & -                                                      & -                                                      & -                                                      & -                                                      & -                                                      & -                                                     & -                                                      & \begin{tabular}[c]{@{}c@{}}3.32  \tiny (1.15)\small \end{tabular} & \begin{tabular}[c]{@{}c@{}}\textbf{4.12 } \tiny (0.9)\small \end{tabular} \\ \hline

\end{tabular}
}
\label{table:voice_evaluation}
\end{table}

\subsection{Video clips}
\begin{figure}[h]
  \includegraphics[width=\linewidth]{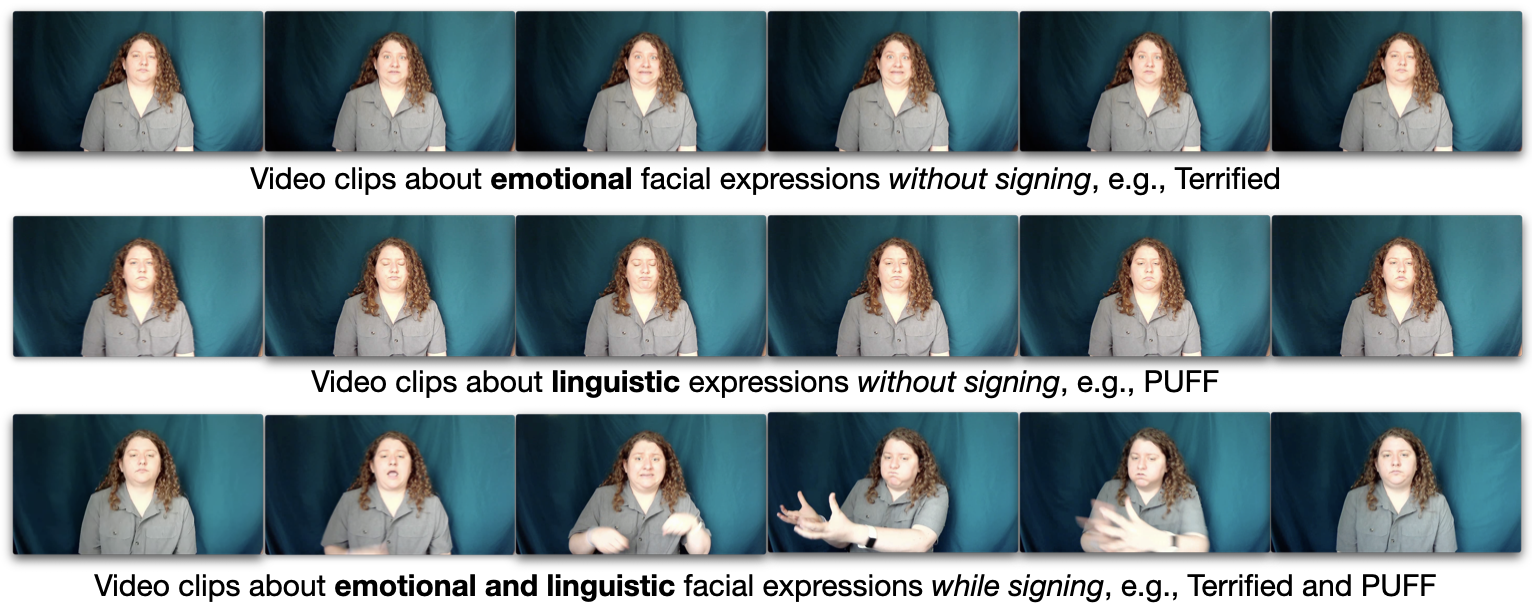}
  \caption{Video Clips for Survey}
  \Description{}
  \label{fig:videoclips}
\end{figure}

The facial expressions used in the study were generated by a deaf sign model (female, 37 years old). None of the sign model was later found to be familiar to the subjects. The sign model was videotaped wearing bright clothing against a dark green background (See Fig\ref {fig:videoclips}
). During the videotaping, the sign model was instructed to produce multiple facial expressions, including 7 linguistic markers and 7 emotional expressions, either with signing sentences or solely through facial expressions. The eye gaze was generally straight ahead but occasionally directed to the side or downward. However, the direction of gaze was balanced across conditions. In total, 102 video clips were created for the survey, featuring:
\begin{itemize} 
\item 7 video clips about emotional facial expressions without signing; no meanings of signs are provided.
\item 7 video clips about linguistic facial expressions without signing; no meanings of signs are provided.
\item 98 video clips about multiple facial expressions while signing with each sign's meaning presented alongside one of three translation types: subtitles, neutral voice, or emotional voices. 
\end{itemize}

For the translation information, we created three types of videos: 1) videos with no voice (only subtitles), 2) videos with a neutral voice, and 3) videos with a corresponding emotional tone that matched the emotional facial expressions. The translation (e.g., subtitles or speech) is displayed for 2 seconds after the signing model finishes. The voice or subtitle appeared immediately after the model finished signing, as we aimed to emulate the translation system. Each video clip, including translation information, lasts approximately 6 seconds. Survey participants were expected to first observe the signer's signing and facial expressions and then view or listen to the translated English.

\subsection{Data Collection}
\subsubsection{Survey Procedure}
Non-ASL hearing respondents were asked to watch 102 video stimuli and answer the following 4 questions for each videos: the meaning of sign and emotional status with the difficulty of answering each questions. To simulate realistic conversation, where hearing individuals could listen and watch the signer’s signs only once, we asked participants to watch each video once and then answer the questions. The order of the videos was randomized. The 45 participants were randomly assigned to one of three conditions:
\begin{itemize}
    \item Condition A: Videos with no voice (only subtitles)
    \item Condition B: Videos with a neutral voice
    \item Condition C: Videos with a corresponding emotional tone that matched the emotional facial expressions
\end{itemize}

The survey took approximately 45 to 60 minutes to complete. Participants received a \$15 Amazon gift card or cash as compensation. Additionally, DHH participants completed a survey where they ranked emotions by importance, provided feedback, and suggested features for developing an emotional voice component for ASL translation systems. The survey took less than 30 minutes, and participants were compensated with a \$10 Amazon gift card or cash.

\subsubsection{Data Analysis}
We collected 45 responses from hearing participants (age range: 18 to 43, mean: 21.93, SD: 4.74; male: 14, female: 31) across three conditions, with 15 responses for each condition. For the analysis, we used 43 responses, as two responses was not reliable for Condition A due to identical answers on all questions. To answer our four research questions, we will analyze the participants' performance in emotion perception, the accuracy of understanding sign meanings, and the mental effort required to answer the two questions across three conditions. Additionally, we will report which emotions were deemed most important based on the rankings and identify further emotions that should be included in the system, along with features for the emotional voice component. 

\begin{table}[h]
\caption{DHH Participants' Demographic Information: 'Advanced' - Proficiency in familiar and unfamiliar topics, fluent and spontaneous expressions without hesitation, discussion in formal and informal settings, excellent comprehension across a wide spectrum of topics and 'Intermediate' - Having social conversations, discussing and describing life experiences/events, more than basic level of comprehension}
\scalebox{0.75}{
\begin{tabular}{c|c|c|c|c|c}
\hline
\textbf{Participant} & \textbf{Gender} & \textbf{Age} & \textbf{Hearing status} & \textbf{ASL Experience} & \textbf{Frequency of Interaction with non-ASL users} \\ \hline
DHH01                & Women           & 27           & Deaf                                  & Advanced                                & Multiple times per day                               \\ \hline
DHH02                & Men             & 25           & deaf                                  & Intermediate                            & Multiple times per day                               \\ \hline
DHH03                & Others          & 26           & Deaf                                  & Intermediate                            & A few times per week                                 \\ \hline
DHH04                & Women           & 29           & Hard of hearing                       & Intermediate                            & Multiple times per day                               \\ \hline
DHH05                & Others          & 22           & Deaf                                  & Advanced                                & Multiple times per day                               \\ \hline
DHH06                & Men             & 17           & Deaf                                  & Advanced                                & Once per day                                         \\ \hline
\end{tabular}
}
\label{table:demoinfo}
\end{table}

Additionally, we gathered six responses from DHH participants (ages 17 to 29, mean age: 24.29, SD: 3.9; 2 males, 2 females, 2 identifying as other, See Fig. Table \ref{table:demoinfo}) regarding the ranking of emotions, suggestions for additional emotions, and features for the system. We will analyze these responses in comparison to those from hearing participants to extract insights on designing emotional voices for ASL translation systems tailored for users.

\section{Result}

In this section, we address each research question based on data analysis. First, we examine emotional facial expressions in two contexts: reading facial expressions alone versus reading both facial expressions and the meaning of the sign. Second, we investigate linguistic markers in detail, particularly how non-ASL hearing individuals perceive linguistic expressions. Third, we explore how hearing participants process multiple facial expressions to recognize emotion and the role of emotional voice in this context. Lastly, we examine the key characteristics of emotional voice. 

\subsection{RQ1: How do non-ASL hearing individuals recognize signers’ emotions when observing both emotional facial expressions and translation information in a communication scenario?}

In this section, we present three main findings regarding emotion perception performance and the mental effort involved in understanding emotion and translation information: (1) Decreased Emotion Perception while Signing, (2) Misinterpretation as 'Neutral' Emotion, and (3) Effect of Emotional Voice. Here, we analyzed the data where there were no linguistic markers, focusing on a single facial expression for emotion, distinguishing between emotion expressed alone and emotion expressed while signing.

\begin{figure}[h]
    \includegraphics[width=1\linewidth]{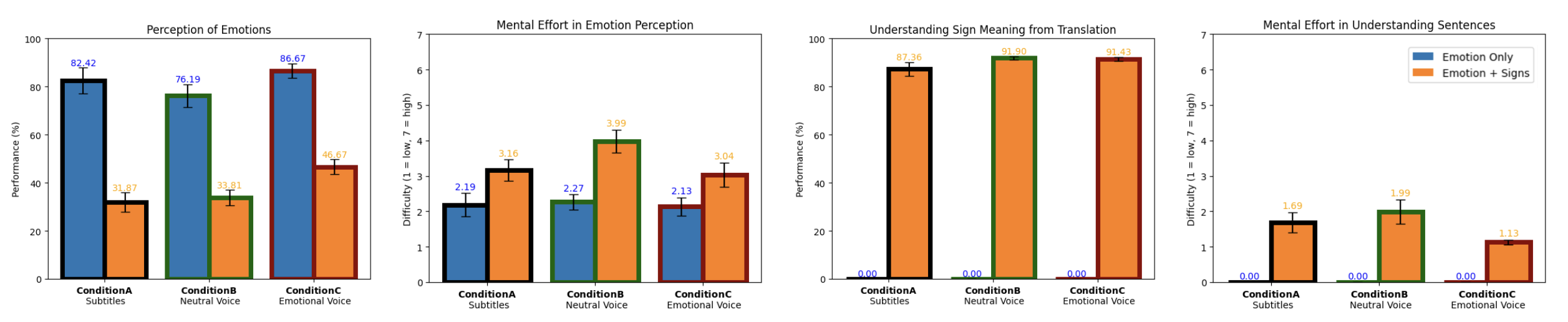}
    \caption{Results comparing reading facial expressions alone (marked in blue) versus reading both facial expressions and the meanings of signs (marked in orange): (A) Emotion Perception, (B) Mental Effort in Emotion Perception, (C) Understanding Meaning from Translation, (D) Mental Effort in Understanding Sentences}
    \label{fig:single_emotion_result_perception}
\end{figure}

\subsubsection{Decreased Emotion Perception while Signing}
We hypothesize that participants have difficulty recognizing the signer's emotion when both observing emotional facial expressions and understanding the meaning of signs through translation, as they are required to process both pieces of information almost simultaneously. We found that \textbf{participants had a significantly lower perception of a signer’s emotions when they understood translated signs via text or speech in addition to emotional facial expressions} compared to when they were only reading the emotional facial expressions (see Fig \ref{fig:single_emotion_result_perception}). This pattern was consistent across all conditions, indicating a statistically significant difference based on a paired t-test: Condition A: p < 0.001, t(12)  = 10.970, d = 2.95; Condition B: p < 0.001, t(14)  = 7.572, d = 2.71; Condition C: p < 0.001, t(14)  = 13.601, d = 3.457. 

It is possible that the added translation (i.e., subtitles or voices) may have introduced cognitive overload or distraction for non-ASL users. Due to their lack of familiarity with sign language, interpreting both the visual and audio elements of translation alongside the signer’s facial expressions required more mental effort. Fig \ref{fig:single_emotion_result_perception} (B) shows that \textbf{mental effort was higher when observing both translation and emotional facial expressions across all conditions} (Condition A: p = 0.012, t(12)  = 2.9797, d = 0.851; Condition B: p < 0.001, t(14)  = 7.2486, d = 1.628; Condition C: p = 0.005, t(14)  = 3.255, d = 0.767), indicating that this increased effort may detract from their ability to fully perceive and interpret the emotional content.

\subsubsection{Misinterpretation as "neutral" Emotion during Signing}
\begin{figure}[h]
    \includegraphics[width=1\linewidth]{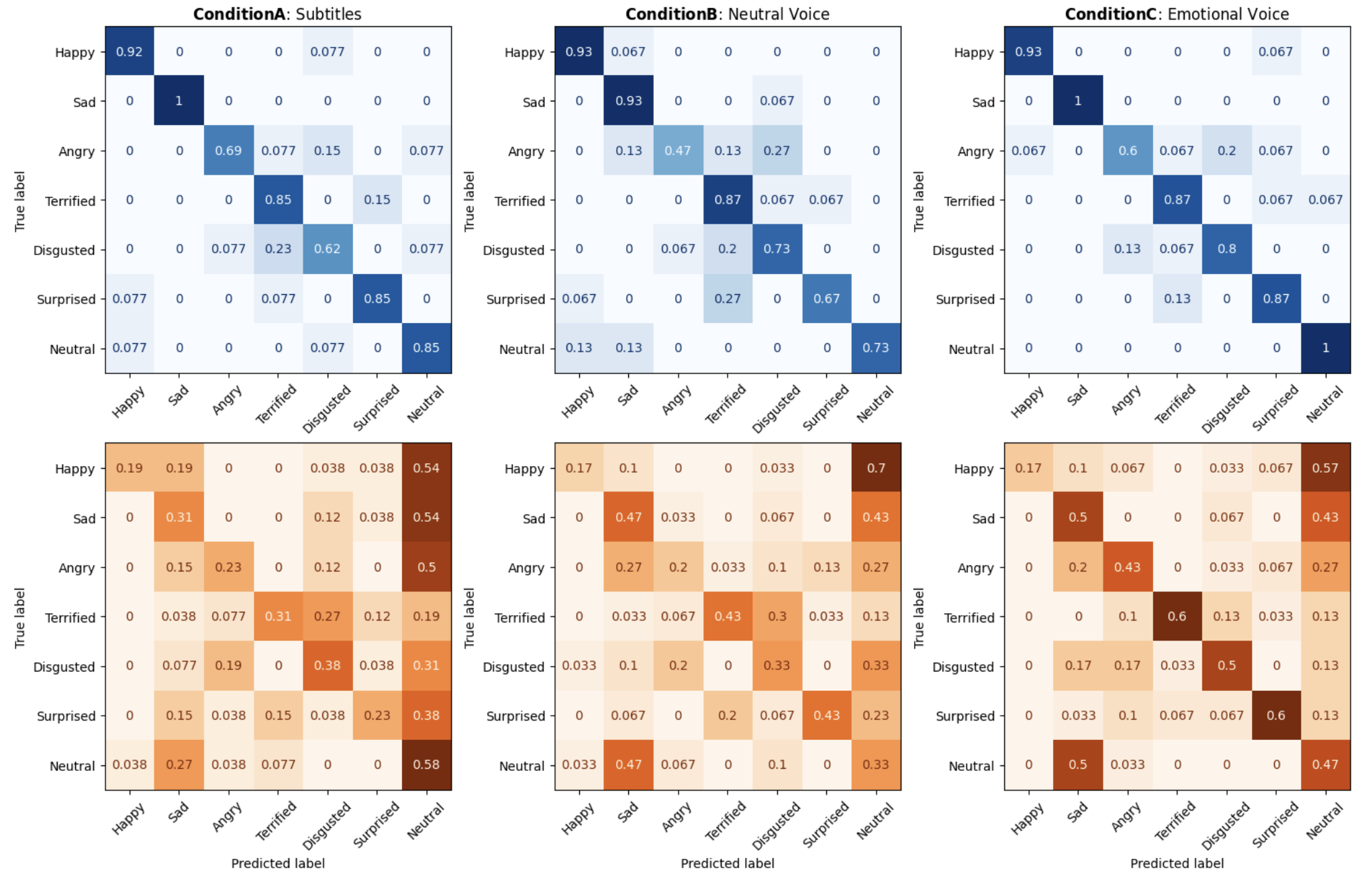}
    \caption{Confusion Matrix over Three Conditions: displaying the performance of emotion perception when observing emotional facial expressions alone (marked in blue) compared to observing emotional facial expressions during signing (marked in orange).}
    \label{fig:Confusion Matrix Emotion only}
    \Description{}
\end{figure}
The main reason for the drop in performance when observing both translations and facial expressions is that \textbf{participants largely misperceived the signer’s emotions as "neutral"} (see Fig. \ref{fig:Confusion Matrix Emotion only}). This misperception was likely due to the brevity of the emotional facial expressions and the overall neutral appearance during signing. Typically, emotional facial expressions occurred within approximately 500 to 1000 ms in the 6-second video clips. In contrast, the videos where emotions were only expressed lasted about 2 to 3 seconds, indicating that the duration of emotional expression while signing was shorter, which likely negatively impacted participants' ability to accurately perceive the signer’s emotions. 

Surprisingly, \textbf{the perception of happy emotions had low performance, often being perceived as "neutral"}, as the signer’s happy facial expressions were very quick and featured only a faint smile, which was not distinct from neutral facial expression during signing. Additionally, we found that participants perceived the signer’s neutral facial expression as "sad", which may be signer-dependent. P28 commented that \textit{"...I kept thinking "she looks bored" during the survey and ended up going with "neutral" or "sad" on a lot of them..."}

\subsubsection{Effect of Emotional Voice}
We hypothesize that corresponding emotional voices accompanying facial expressions in translation help non-ASL hearing participants perceive emotions more effectively. A one-way ANOVA was conducted to compare the performance on the perception of the signer’s emotions across three Condition A, B, and C. There was a significant difference in performance, F(2,40)=5.747375 , p=0.006396, with a moderate effect size ( \(n^2\) =0.2232). Post hoc comparisons using Tukey’s HSD test indicated that participants in Condition C (M = 46.67\%, SD = 11.726) scored significantly higher than those in Condition A (p=0.0108, M = 31.87\%, SD = 14.46\%) and Condition B (p=0.0233, M = 33.81\%, SD = 12.21\%), suggesting that \textbf{the emotional voice helped enhance perception of the signer's emotions.} However, there were no significant differences between Condition A and Condition B (p=0.9153). 

As shown in Fig \ref{fig:Confusion Matrix Emotion only}, certain emotions, such as angry, terrified, disgusted, and surprised, were perceived more accurately, likely because audio cues might help hearing participants make their choices, even if they missed the visual cues. On the other hand, happy emotions did not benefit from audio cues, as we believe that the happy emotional voice may not have been as clear compared to the others. We will further discuss the relationship between emotional voice clarity and our performance in Section \ref{emotionalvoice_discussion}.

For mental effort, the results were not statistically significant, F(2,40)=2.619315  , p=0.085312, with a moderate effect size (\(n^2\) =0.115799). Post hoc comparisons using Tukey’s HSD test indicated that the difference between participants in Condition B and Condition C approached significance but was not statistically significant (p=0.0853). This suggests that while there may be a trend towards a difference, indicating that\textbf{ the emotional voice may help reduce the mental effort required to recognize emotions}.

For the accuracy of understanding signs, there was no significant difference across the conditions F(2,40)=2.339168  , p=0.109461, with a small effect size ( \(n^2\)=0.10471). All conditions achieved over 87.37\% (SD = 10.16\%), 91.90\% (SD = 2.51\%), and 91.43\% (2.95\%). However, regarding mental effort, even though all conditions indicated that no difference (F(2,40)=2.914319  , p=0.0658, with a small effect size ( \(n^2\)=0.1271)) and participants found it easy to understand the signs, \textbf{there was a trend suggesting that the emotional voice (Condition C, M = 1.13, SD = 0.255) was more helpful (p=0.056) than the neutral voice (Condition B, M = 1.99, SD = 1.335) in understanding signs}, based on the results from post hoc comparisons using Tukey’s HSD test.

\subsection{RQ2: How do non-ASL hearing individuals recognize signers’ emotions when observing linguistic facial expressions of a signer?}

In this section, we present three main findings regarding linguistic facial expression: (1) Interpretation of Linguistic Markers as Emotions, (2) Mental Effort on Linguistic Facial Expressions, and (3) Effect of Emotional Voice against Linguistic Facial Expressions. Here, we analyzed data in which the signer’s emotional status was always 'neutral' along with linguistic facial expressions. Thus, we explore how non-hearing individuals perceive the signer’s emotions based on linguistic facial expressions, even when the signer’s emotional state is neutral.

\subsubsection{Interpretation of Linguistic Markers as Emotions}

\begin{figure}
    \includegraphics[width=1\linewidth]{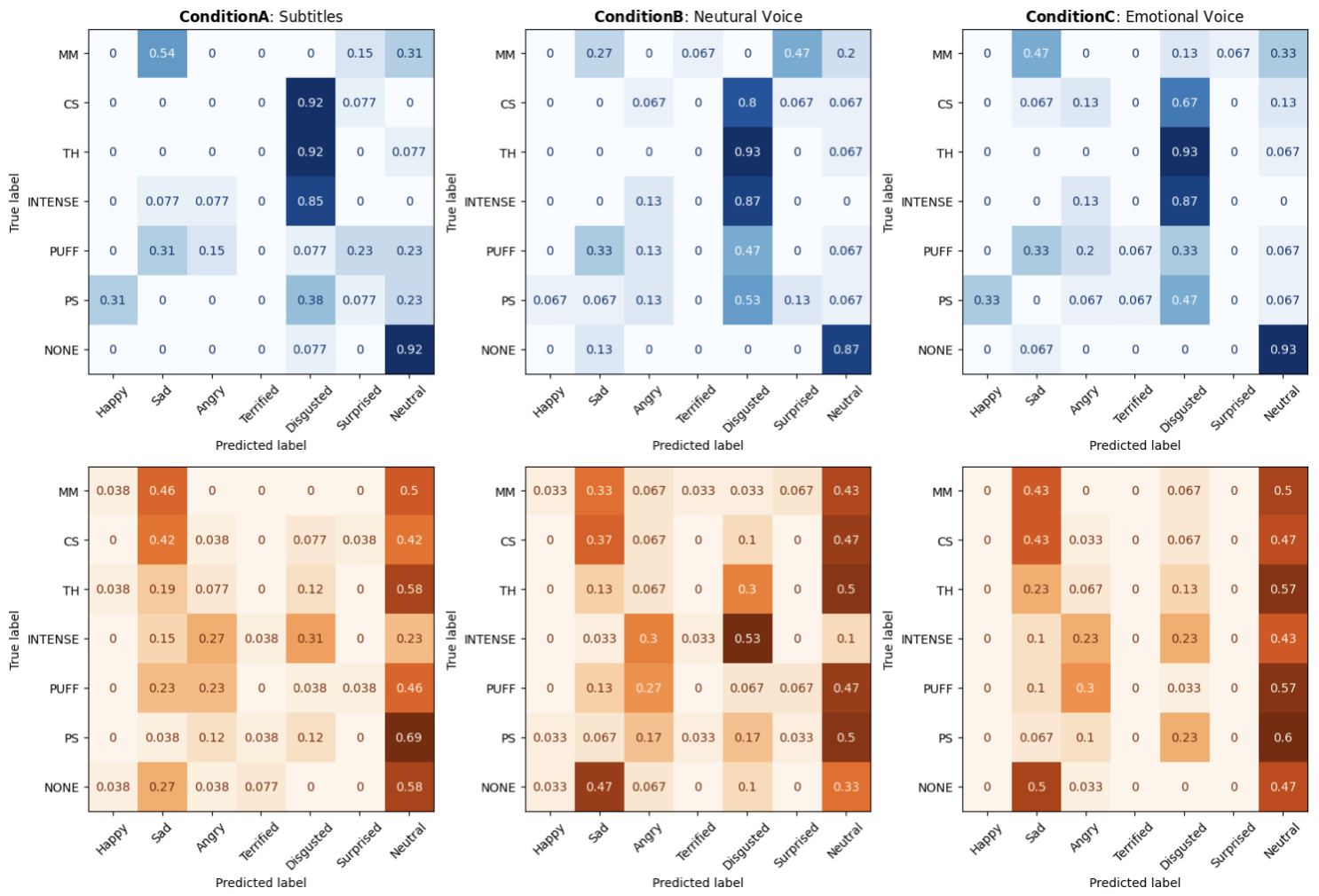}
    \caption{Confusion Matrix over Three Conditions: Displaying the ratio of emotion perception when observing linguistic facial expressions alone (marked in blue) compared to observing linguistic facial expressions during signing (marked in orange).}
    \label{fig:Confusion Matrix_lingustic}
    \Description{}
\end{figure}

We hypothesize that linguistic facial expressions of a signer may be interpreted as emotional expressions by non-hearing individuals, as they may not recognize these facial expressions as linguistic markers. As shown in Fig \ref{fig:Confusion Matrix_lingustic}, the results show that \textbf{when observing linguistic facial expressions without signing, linguistic markers were commonly misinterpreted as negative emotions such as 'disgust' or 'sad.'} The linguistic markers CS, TH, INTENSE, and PS were identified as 'disgust' at rates of 79\%, 93\%, 86\%, and 47\%, respectively. These linguistic markers tend to produce  a degree of frowning in the eyebrows, in facial expressions that are likely perceived as disgusted. Also, MM and PUFF use the lips and cheeks to protrude, making them resemble a 'sad' facial expression in the signer. On the other hands, PS was only perceived as positive emotion such as 'happy' with a rate of 27\%, as it involves tightening the lips, which can be interpreted as a smile. 

However, the results shifted when linguistic facial expressions were observed during signing. Participants tended to select 'neutral' emotions even when linguistic facial expressions were present. Overall three Condition A, B, and C, it achieved the accuracy of 49.45\% (SD = 26.64\%), 40.0\% (SD = 18.86\%), and 51.43\% (SD = 23.09\%), respectively. \textbf{This outcome does not indicate that non-ASL hearing participants recognized and intentionally disregarded linguistic markers.} Rather, it may be due to the brevity of linguistic expressions—lasting only about 500 to 1000 ms in a 6-second video clip—and the subtlety of certain linguistic markers, such as CS and PS, during signing. Additionally, the increased rate of 'sad' responses may be influenced by signer-dependent factors, as the signer’s neutral facial expression was often perceived as 'sad,' contributing to this pattern.

\subsubsection{Mental Effort on Linguistic Facial Expressions}

\begin{figure}
    \includegraphics[width=1\linewidth]{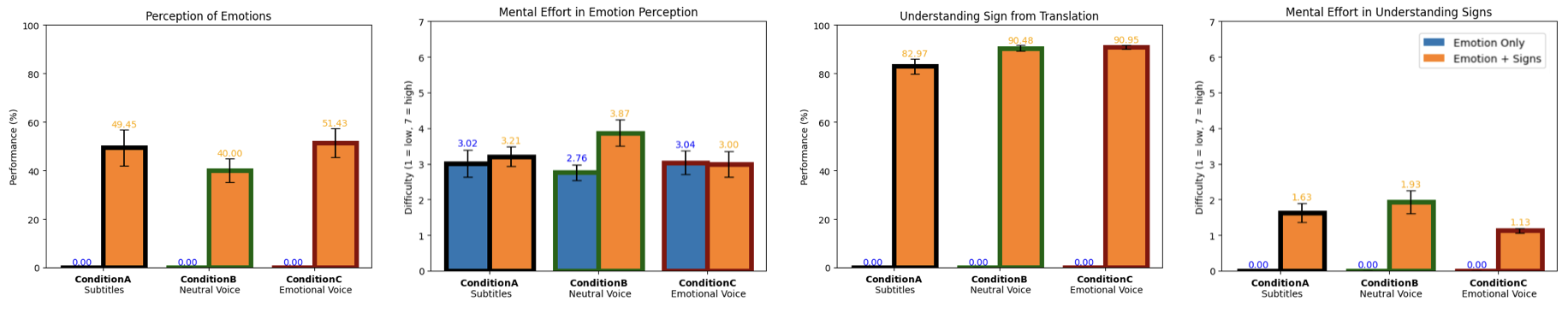}
    \caption{Results comparing reading only linguistic facial expressions without signing (marked in blue) versus reading linguistic facial expressions while signing (marked in orange): (A) Emotion Perception, (B) Mental Effort in
Emotion Perception, (C) Understanding Meaning from Translation, (D) Mental Effort in Understanding
Sentences}
    \label{fig:single_linguistic_result_perception}
\end{figure}

We hypothesize that non-ASL hearing individuals experience increased confusion and mental effort when trying to perceive a signer’s emotions, especially since linguistic facial expressions are not related to emotions at all. As shown in Fig \ref{fig:single_linguistic_result_perception} (B), the results showed that \textbf{reading linguistic facial expressions as emotions requires a higher mental load }(M = 2.936, SD = 1.199) compared to reading emotional facial expressions (M = 2.196, SD = 0.993) (p < 0.001, t(42) = 5.988, d = 0.672). This is primarily because linguistic markers do not convey emotions, leading to participant confusion and increased mental effort. For example, linguistic facial expressions, such as PUFF, can be interpreted as various emotions, including sadness, anger, disgust, surprise, and neutrality.

We expected that mental load would arise due to the ambiguity of linguistic facial expressions for emotion and the need to process these to understand the translation. Interestingly, however, \textbf{mental effort did not change when observing both linguistic facial expressions and translation during signing }(see Fig. 5). This may be because the mental effort stems more from the simultaneous processing of both the emotions and the meanings of the signs provided by translation, rather than from interpreting the facial expressions themselves for emotion. This is supported by the lack of difference in mental effort during signing when comparing observations of only linguistic facial expressions and only emotional facial expressions across all conditions (See Fig \ref{fig:single_emotion_result_perception} (B) and Fig \ref{fig:single_linguistic_result_perception} (B) (Condition A: p = 0.6771, t(12) = 0.42664, d = 0.0466; Condition B: p = 0.547, t(14) = 0.6158, d = 0.083; Condition C: p = 0.777, t(14) = 0.28848, d = 0.02758).
\subsubsection{Effect of Emotional Voice against Linguistic Facial Expressions}

We hypothesized that an emotional voice would assist non-ASL hearing participants in accurately interpreting the signer’s emotions, even in the presence of linguistic markers. This is because participants may rely on audio cues, even if the facial expression does not align with the emotion. As shown in Fig \ref{fig:single_linguistic_result_perception} (A), the results suggest that \textbf{an emotional voice (Condition C) matching the signer’s emotional state supports non-ASL hearing participants in accurately perceiving the intended emotion (i.e., 'neutral'), even when linguistic facial expressions are present. } We also observed that participants in Condition B (neutral voice) tended to rely on visual cues, such as facial expressions, despite the presence of a neutral voice. This likely occurred because participants perceived the neutral voice as a default without emotional information, often disregarding it when identifying the signer’s emotion. For example, as shown in Fig \ref{fig:Confusion Matrix_lingustic}, with the INTENSE linguistic marker, participants in Condition B likely relied primarily on visual cues, with only 10\% selecting 'neutral.' In Condition C, however, audio cues had a stronger influence, leading the participants to select 'neutral' 43\% of the time.


\subsection{RQ3: How do non-ASL hearing individuals recognize signers’ emotions when observing multiple facial expressions such as both emotional and linguistic facial expression simultaneously during signing?}

We have so far examined emotional and linguistic facial expressions separately to understand how non-ASL hearing individuals interpret each. However, in real-world scenarios, signers often use both types simultaneously. We further explored how participants perceive emotions when both emotional and linguistic facial expressions are present and how emotional voice aids in accurate emotion recognition.

\subsubsection{Confusion on Emotions from Multiple Facial Expressions} 
\begin{table}[h]
\caption{Top 10 Lowest Emotion Perception Performances, with mean and standard deviation in parentheses: Most of these instances occur when both emotional and linguistic facial expressions are present. \textit{Italic text }represents a single facial expression, whether emotional or linguistic.}
\scalebox{0.9}{
\begin{tabular}{cccccc}
\hline
\multicolumn{6}{c}{\textbf{Top 10 Lowest Emotion Perception Performances}}                                                                                                                                              \\ \hline
\multicolumn{2}{c|}{Condition A: Subtitles}                                & \multicolumn{2}{c|}{Condition B: Neutral Voice}                          & \multicolumn{2}{c}{Condition C: Emotional Voice}      \\ \hline
\multicolumn{1}{c|}{Terrified\_PUFF}    & \multicolumn{1}{c|}{0 (0)}       & \multicolumn{1}{c|}{Terrified\_PUFF}  & \multicolumn{1}{c|}{0.03 (0.18)} & \multicolumn{1}{c|}{Happy\_TH}          & 0.1 (0.31)  \\ \hline
\multicolumn{1}{c|}{Surprised\_TH}      & \multicolumn{1}{c|}{0.12 (0.33)} & \multicolumn{1}{c|}{\textit{Neutral\_INTENSE}} & \multicolumn{1}{c|}{0.1 (0.31)}  & \multicolumn{1}{c|}{Disgusted\_CS}      & 0.13 (0.35) \\ \hline
\multicolumn{1}{c|}{Happy\_TH}          & \multicolumn{1}{c|}{0.12 (0.33)} & \multicolumn{1}{c|}{Angry\_TH}        & \multicolumn{1}{c|}{0.13 (0.35)} & \multicolumn{1}{c|}{\textit{Happy\_NONE}}        & 0.17 (0.38) \\ \hline
\multicolumn{1}{c|}{Happy\_INTENSE}     & \multicolumn{1}{c|}{0.12 (0.33)} & \multicolumn{1}{c|}{\textit{Happy\_NONE}}      & \multicolumn{1}{c|}{0.17 (0.38)} & \multicolumn{1}{c|}{Terrified\_PUFF}    & 0.2 (0.41)  \\ \hline
\multicolumn{1}{c|}{Surprised\_INTENSE} & \multicolumn{1}{c|}{0.15 (0.37)} & \multicolumn{1}{c|}{Sad\_TH}          & \multicolumn{1}{c|}{0.17 (0.38)} & \multicolumn{1}{c|}{Happy\_INTENSE}     & 0.2 (0.41)  \\ \hline
\multicolumn{1}{c|}{Angry\_TH}          & \multicolumn{1}{c|}{0.15 (0.37)} & \multicolumn{1}{c|}{Disgusted\_CS}    & \multicolumn{1}{c|}{0.17 (0.38)} & \multicolumn{1}{c|}{Disgusted\_TH}      & 0.3 (0.47)  \\ \hline
\multicolumn{1}{c|}{Angry\_MM}          & \multicolumn{1}{c|}{0.15 (0.37)} & \multicolumn{1}{c|}{Surprised\_TH}    & \multicolumn{1}{c|}{0.2 (0.41)}  & \multicolumn{1}{c|}{Surprised\_PUFF}    & 0.3 (0.47)  \\ \hline
\multicolumn{1}{c|}{Terrified\_CS}      & \multicolumn{1}{c|}{0.19 (0.4)}  & \multicolumn{1}{c|}{\textit{Angry\_NONE}}      & \multicolumn{1}{c|}{0.2 (0.41)}  & \multicolumn{1}{c|}{Terrified\_INTENSE} & 0.3 (0.47)  \\ \hline
\multicolumn{1}{c|}{Surprised\_PS}      & \multicolumn{1}{c|}{0.19 (0.4)}  & \multicolumn{1}{c|}{Happy\_TH}        & \multicolumn{1}{c|}{0.2 (0.41)}  & \multicolumn{1}{c|}{Angry\_TH}          & 0.3 (0.47)  \\ \hline
\multicolumn{1}{c|}{\textit{Happy\_NONE}}        & \multicolumn{1}{c|}{0.19 (0.4)}  & \multicolumn{1}{c|}{Terrified\_TH}    & \multicolumn{1}{c|}{0.2 (0.41)}  & \multicolumn{1}{c|}{Sad\_TH}            & 0.37 (0.49) \\ \hline
\end{tabular}
}
\label{table:top10-accuracy}
\end{table}

\begin{table}[h]
\caption{Top 10 Highest Mental Effort Scores (1 = low, 7 = high), with mean and standard deviation in parentheses: Most of these instances occur when both emotional and linguistic facial expressions are present. Italic text represents a single facial expression, whether emotional or linguistic.}
\scalebox{0.9}{
\begin{tabular}{cccccc}
\hline
\multicolumn{6}{c}{\textbf{Top-10 Highest Mental Effort Score}}                                                                       \\ \hline
\multicolumn{2}{c|}{Condition A: Subtitles}                                & \multicolumn{2}{c|}{Condition B: Neutral Voice}                          & \multicolumn{2}{c}{Condition C: Emotional Voice}      \\ \hline
\multicolumn{1}{c|}{Happy\_INTENSE}     & \multicolumn{1}{c|}{4.31 (1.67)} & \multicolumn{1}{c|}{Happy\_MM}        & \multicolumn{1}{c|}{4.63 (1.92)} & \multicolumn{1}{c|}{Angry\_MM}          & 3.77 (1.87) \\ \hline
\multicolumn{1}{c|}{Terrified\_MM}      & \multicolumn{1}{c|}{4.19 (1.72)} & \multicolumn{1}{c|}{Angry\_CS}        & \multicolumn{1}{c|}{4.43 (2.03)} & \multicolumn{1}{c|}{Terrified\_PUFF}    & 3.7 (1.86)  \\ \hline
\multicolumn{1}{c|}{Surprised\_INTENSE} & \multicolumn{1}{c|}{4.15 (1.52)} & \multicolumn{1}{c|}{Happy\_PUFF}      & \multicolumn{1}{c|}{4.4 (1.75)}  & \multicolumn{1}{c|}{Terrified\_PS}      & 3.57 (2.06) \\ \hline
\multicolumn{1}{c|}{Terrified\_TH}      & \multicolumn{1}{c|}{4.12 (1.82)} & \multicolumn{1}{c|}{Surprised\_CS}    & \multicolumn{1}{c|}{4.37 (1.92)} & \multicolumn{1}{c|}{Disgusted\_INTENSE} & 3.53 (1.85) \\ \hline
\multicolumn{1}{c|}{Happy\_PS}          & \multicolumn{1}{c|}{4.12 (1.73)} & \multicolumn{1}{c|}{Sad\_MM}          & \multicolumn{1}{c|}{4.37 (1.79)} & \multicolumn{1}{c|}{Terrified\_MM}      & 3.5 (2.05)  \\ \hline
\multicolumn{1}{c|}{Happy\_PUFF}        & \multicolumn{1}{c|}{3.96 (2.05)} & \multicolumn{1}{c|}{\textit{Disgusted\_NONE}}  & \multicolumn{1}{c|}{4.27 (1.91)} & \multicolumn{1}{c|}{Disgusted\_TH}      & 3.5 (1.83)  \\ \hline
\multicolumn{1}{c|}{Surprised\_PS}      & \multicolumn{1}{c|}{3.96 (1.99)} & \multicolumn{1}{c|}{Happy\_TH}        & \multicolumn{1}{c|}{4.23 (1.7)}  & \multicolumn{1}{c|}{Angry\_CS}          & 3.4 (1.87)  \\ \hline
\multicolumn{1}{c|}{Sad\_INTENSE}       & \multicolumn{1}{c|}{3.88 (1.84)} & \multicolumn{1}{c|}{\textit{Happy\_NONE}}      & \multicolumn{1}{c|}{4.23 (1.61)} & \multicolumn{1}{c|}{Terrified\_INTENSE} & 3.37 (1.92) \\ \hline
\multicolumn{1}{c|}{Disgusted\_PS}      & \multicolumn{1}{c|}{3.85 (1.99)} & \multicolumn{1}{c|}{\textit{Neutral\_INTENSE}} & \multicolumn{1}{c|}{4.17 (1.82)} & \multicolumn{1}{c|}{\textit{Disgusted\_NONE}}    & 3.37 (2.06) \\ \hline
\multicolumn{1}{c|}{Angry\_PS}          & \multicolumn{1}{c|}{3.85 (1.74)} & \multicolumn{1}{c|}{Disgusted\_CS}    & \multicolumn{1}{c|}{4.17 (1.74)} & \multicolumn{1}{c|}{Disgusted\_PUFF}    & 3.37 (2.08) \\ \hline
\end{tabular}
}
\label{table:top10-mental}
\end{table}

We hypothesized that multiple facial expressions may lead to confusion in identifying the signer’s emotions. For instance, if a happy facial expression is presented alongside an INTENSE linguistic marker, non-ASL hearing individuals may experience confusion between emotions such as happiness and disgust. This confusion may intensify when the audio does not align with the visual expressions, creating cognitive dissonance and increasing mental effort in recognizing emotions. Results indicate that t\textbf{he top-10 lowest emotion perception accuracy (See. Table \ref{table:top10-accuracy}) and the top-10 highest mental effort scores (See. Table \ref{table:top10-mental}) predominantly occurred when multiple facial expressions were present.} This suggests that \textbf{conflicting facial expressions can mislead participants in recognizing the signer’s emotional state.} In particular, the combined contradictory expression 'Terrified\_PUFF' was interpreted as 'angry' by participants, with misinterpretation rates of 61\%, 73\%, and 51\% across Conditions A, B, and C, respectively. We also found that the linguistic marker, TH, was the marker that most frequently caused confusion across all three conditions. Participants interpreted TH as “disgusted” with over 92\% ratio in all conditions (See Fig \ref{fig:Confusion Matrix_lingustic}). Consequently, the lowest-performing cases in emotion perception frequently included instances of TH.


\subsubsection{Effect of Emotional Voice against Multiple Facial Expressions}

\begin{figure}[h]
    \includegraphics[width=1\linewidth]{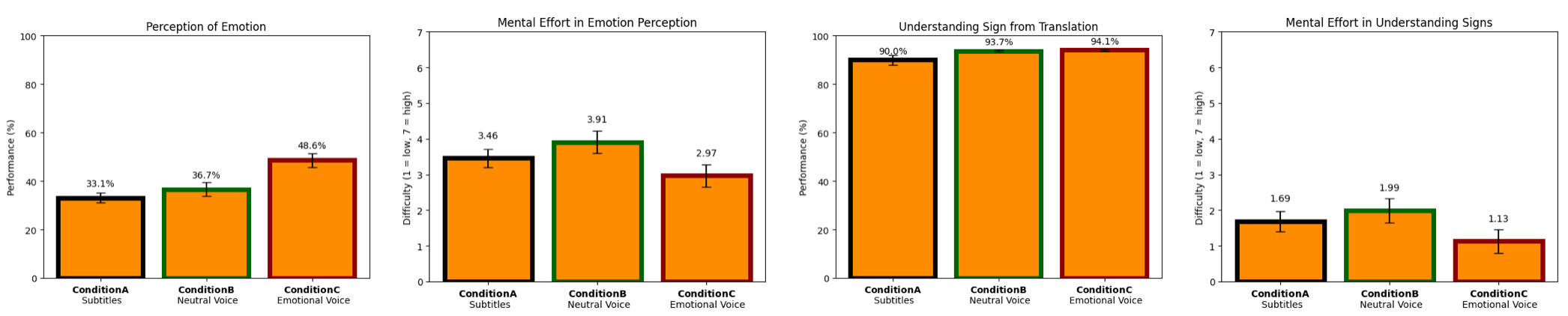}
    \caption{Results when reading both emotional and linguistic facial expressions while signing: (A) Emotion Perception, (B) Mental Effort in
Emotion Perception, (C) Understanding Meaning from Translation, (D) Mental Effort in Understanding
Sings}
    \label{fig:multiple_facial_1}
\end{figure}

Overall, non-ASL hearing participants achieved emotion perception when observing multiple facial expressions with a rates of 33.1\% (SD = 7.8\%) in Condition A, 36.7\% (SD = 11.28\%) in Condition B, and 48.6\% (SD = 11.50\%) in Condition C (See. Fig \ref{fig:multiple_facial_1}). Performance varied significantly among the conditions, F(2,40) = 8.67, p = 0.0007, with a moderate effect size (  \(n^2\) = 0.30). Post hoc comparisons using Tukey’s HSD test revealed that participants in Condition C (emotional voice) scored significantly higher than those in Condition A (subtitles) (p = 0.001) and Condition B (neutral voice) (p = 0.0089), indicating that \textbf{emotional voice significantly enhanced emotion perception even when multiple facial expressions were present}. However, no significant difference was observed between Condition A and Condition B (p = 0.648).

\begin{figure}[h]
    \includegraphics[width=1\linewidth]{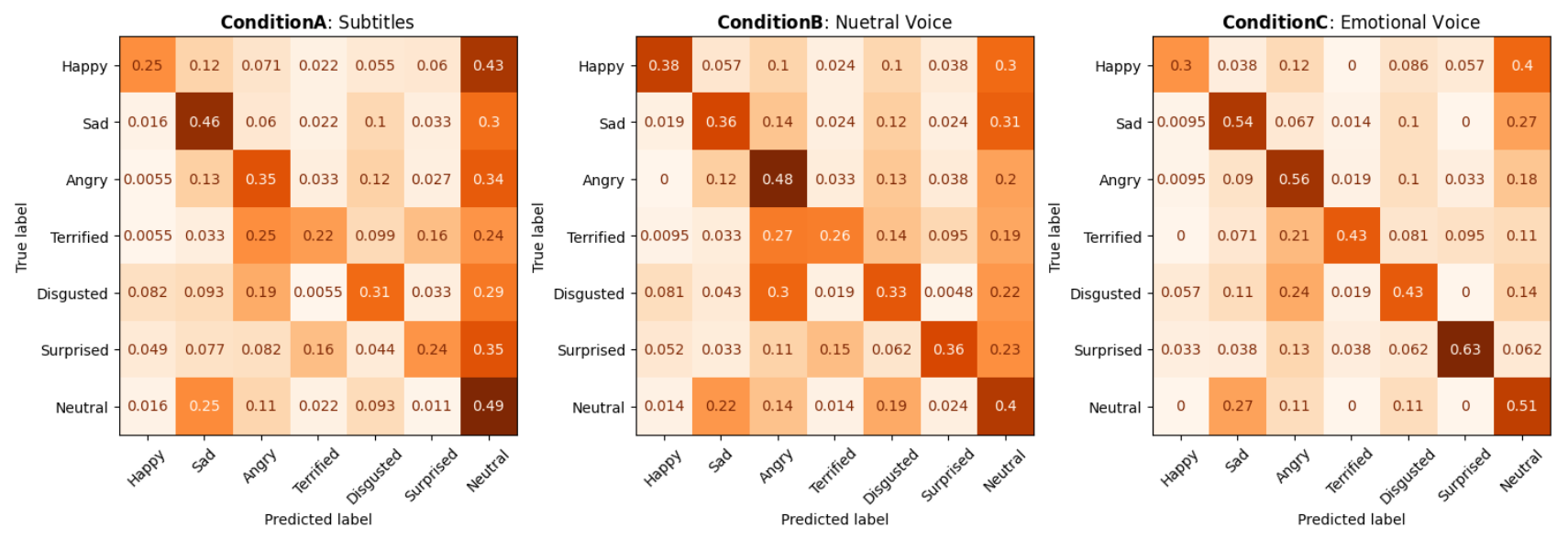}
    \caption{Confusion Matrix Across Three Conditions: Displaying the performance of emotion perception when observing both emotional and linguistic facial expressions during signing.}
    \label{fig:cm_multiple_facial}
\end{figure}

\textbf{An emotional voice aided recognition across all emotions except for "Happy".} Across conditions, "Happy" was notably challenging to identify (See Fig \ref{fig:cm_multiple_facial}), suggesting that the happy emotional voice also did not significantly aid recognition. This may be because the model’s "happy" facial expression was subtle, often resembling a faint smile, which many interpreted as neutral even when paired with linguistic markers. Additionally, the happy emotional voice was rated only moderately effective, scoring 3.6 (SD = 0.86) out of 5 (See Table \ref{table:voice_evaluation}), indicating that the AI-generated happy voice cue might not clearly convey happiness, which limited its effectiveness in supporting happy emotion recognition. 

Conversely, \textbf{a clear emotional voice tone could aid in emotion recognition.} For example, the emotion "terrified" was challenging for participants to identify without audio cues and was frequently misclassified as anger, disgust, or surprise due to its pairing with linguistic markers. However, when a strongly representative "terrified" voice (M = 4.12, SD = 0.9) was provided, accuracy in recognizing "terrified" improved from 22\% (Condition A) and 26\% (Condition B) to 43\% (Condition C), an increase of 40–48\% (See Fig \ref{fig:cm_multiple_facial}). A similar pattern was observed with "surprised," where the addition of the strong emotional voice (M = 4.12, SD = 0.9) increased accuracy from 24\% (Condition A) and 36\% (Condition B) to 63\% (Condition C), an increase of 43–62\% (See Fig \ref{fig:cm_multiple_facial}).

Additionally, the emotional voice could help reduce the mental effort required for emotion perception and understanding signs. This is evident from the trends observed between Conditions B and C, showing differences of mental effort on emotion perception (p = 0.076) (See Fig \ref{fig:multiple_facial_1} (B)) and understanding sign meanings (p = 0.056) (See Fig \ref{fig:multiple_facial_1} (C)) based on post hoc comparisons using Tukey’s HSD test.

\subsubsection{Effect of the Order of Facial Expressions}

In designing the dataset, we considered the order of facial expressions to explore its effect on emotion perception. Order A features linguistic facial expressions first, followed by emotional expressions, while Order B presents emotional expressions first, followed by linguistic markers. We hypothesized that Order A would yield better performance in emotion recognition, as participants tend to rely on the last facial expression to answer questions. Results indicated that \textbf{when emotional expressions were presented last, participants better recognized the signer’s emotions,} particularly in Conditions A (Order A - 36.10\% (SD:7.96\%), Order B - 30.14\% (SD: 9.6\%), p < 0.02, t(12) = 2.670, d = 0.674) and C (Order A - 52.78\% (SD:9.69\%), Order B - 44.48\% (SD:14.76\%), p = 0.0051, t(14) = 3.310, d = 0.664), showing significant differences. No significant difference was observed in Condition B (Order A - 38.09\% (SD:11.69), Order B - 35.23\% (SD: 13.02\%), p = 0.235, t(14) = 1.240, d = 0.235). It is possible that the neutral voice may contradict the last facial emotion, leading to cognitive dissonance that interferes with the perception of emotions.

\subsection{RQ4: How should emotional voice be designed for emotional-voice translation systems to aid communication between hearing and Deaf or Hard of Hearing individuals?}

Based on the results, we found that emotional voice can assist non-ASL hearing individuals in recognizing emotional states. However, the performance is still limited, indicating room for improvement in the design of voice features. Additionally, it is essential to explore the types of emotional voice cues that should be prioritized for user usability.

In this section, we present survey findings on designing emotional voice from a user perspective. Responses were collected from 6 DHH and 43 non-ASL hearing participants, who provided feedback on desired features for emotional voice, ranked emotional voice types by importance, and suggested additional emotional voice options they would like to see included.

\subsubsection{Ranking the Importance of Emotional Voice}
We explored which emotions are most important to users and which are most needed for the system. Participants were asked to rank seven emotions and provide reasoning for their choices.

Overall, there is no significant difference in the importance ranking of emotional voices between hearing individuals and those who are Deaf or Hard of Hearing (DHH) (p > 0.05, Mann-Whitney U-Test). Participants largely based their choices on the frequency of emotion use and the potential effects when emotions are misinterpreted.

\begin{table}[h]
\caption{Ranking the Importance of Emotional Voices}
\scalebox{0.95}{
\begin{tabular}{c|c|c|c|c|c|c|c|c}
\hline
\multicolumn{1}{l|}{}    & \multicolumn{1}{l|}{} & \multicolumn{1}{l|}{happy} & \multicolumn{1}{l|}{sad} & \multicolumn{1}{l|}{angry} & \multicolumn{1}{l|}{terrified} & \multicolumn{1}{l|}{disgust} & \multicolumn{1}{l|}{surprised} & \multicolumn{1}{l}{neutral} \\ \hline
\multirow{3}{*}{DHH}     & Ranking               & \textbf{2}                 & \textbf{4}               & \textbf{1}                 & \textbf{3}                     & \textbf{6}                   & \textbf{5}                     & \textbf{7}                  \\ \cline{2-9} 
                         & Mean                  & 3.14                       & 4.00                     & 1.86                       & 3.43                           & 5.29                         & 4.86                           & 5.43                        \\ \cline{2-9} 
                         & SD                    & (1.77)                     & (1.00)                    & (0.69)                     & (1.72)                         & (1.11)                       & (2.19)                         & (2.70)                      \\ \hline
\multirow{3}{*}{Non-ASL Hearing} & Ranking               & \textbf{4}                 & \textbf{2}               & \textbf{1}                 & \textbf{3}                     & \textbf{6}                   & \textbf{7}                     & \textbf{5}                  \\ \cline{2-9} 
                         & Mean                  & 4.19                       & 3.47                     & 3.16                       & 3.53                           & 4.53                         & 4.79                           & 4.33                        \\ \cline{2-9} 
                         & SD                    & (2.23)                      & (1.82)                   & (1.59)                     & (1.76)                         & (1.93)                       & (1.92)                         & (2.24)                      \\ \hline
\end{tabular}
}
\label{table:Ranking}
\end{table}

\textbf{Anger ranked highest for both DHH and non-ASL hearing participants} (See Table \ref{table:Ranking}). This is because anger is often expressed in various contexts. DHH04 stated, \textit{"Happy is the most popular, and anger is second. These emotions are often used in ASL." }Participants emphasized that addressing negative emotions effectively is crucial; otherwise, it can lead to communication difficulties. As P02 noted, \textit{"Usually, you don't want to misinterpret anger because that would not be good for you, so I put other negative emotions higher."} DHH01 added, \textit{"Anger is one of the most intense emotions to convey."}

\textbf{The emotion of terrified ranked third for both groups.} Participants mentioned that this emotion relates to 'safety' issues and can easily be confused with other emotions, making it vital for clear communication. DHH01 remarked, \textit{"Happy, angry, and terrified are the top important emotions that should be conveyed right away, especially terrified, because it indicates that there’s something wrong that needs to be resolved."} H39 commented, \textit{"I can't really tell terrified from angry or surprised." }H27 noted, \textit{"For me, terrified and surprised both involve raised eyebrows and rounded eyes, so these two are the hardest emotions to differentiate."}

\textbf{We found the neutral emotion interesting, even though it received a low ranking.} This is because a neutral emotional voice is often used and can be difficult to distinguish due to the absence of strong facial expressions. H09 stated, \textit{"Emotions like happiness and disgust will have strong facial expressions. However, I don’t think no facial expressions could always represent neutral."} H34 added, \textit{"Neutral might be hard to define, as it varies by individual."} H25 emphasized, \textit{"Most conversations are neutral, so we don’t want those to be misinterpreted."} 

\subsubsection{Additional Emotion}
Survey responses revealed that \textbf{both non-ASL hearing and DHH participants would like additional emotional voices into translation system to enhance communication clarity}, as they feel that seven emotions are insufficient to fully express the range of human emotions.

\begin{table}[h]
\caption{Additional Emotions suggested by Non-ASL Hearing and DHH Individuals: Number of Mentions by Participants (in parentheses) }
\begin{tabular}{c|c}
\hline
\textbf{Non-ASL Hearing Individual} & \textbf{DHH}\\ \hline
\begin{tabular}[c]{@{}c@{}}anxious (3), disappointed (3), speechless (3), tired (3), bored (2), scared (2), \\ annoyed (1), appreciative (1), confused (1), consoling (1), contentment (1), \\ depressed (1), ecstatic (1), frustrated (1), glad (1), helpless (1), indifference (1), \\ jealous (1), reluctant (1), satisfied (1), shame (1)\end{tabular} & \begin{tabular}[c]{@{}c@{}}frustrated (2)\\ confused (1)\\ impatience (1)\\ shocked (1)\end{tabular} \\ \hline
\end{tabular}
\label{Table:additional emotions}
\end{table}

For\textbf{ hearing participants}, 53\% (23 responses) suggested that additional emotions should be considered, collectively mentioning over 21 distinct emotions (See Table \ref{Table:additional emotions}). For example, P24 said \textit{"Human emotion is a spectrum, so it's hard to cherry-pick individual emotions like this."} Additionally, three participants expressed a desire for more positive emotions (e.g., ecstatic, contentment, glad) as they felt the current range lacked sufficient positive representation. This indicates that hearing participants may prefer a spectrum of positive emotions to better capture subtle nuances in the signer’s emotional state rather than perceiving only a general 'happy' sentiment.
\textit{"I think there should be more positive emotions since there's only happy right now."} P21 said. P22 continued:\textit{"Other emotions should be provided. Happy was the only positive emotion aside from possibly surprised."}

Among\textbf{ DHH participants}, all except one expressed a desire to include additional emotions, believing that the current selection of seven is insufficient for capturing the complexity of emotional expression. They suggested that a broader range of emotions or an option to select from a wider spectrum would be more representative. "\textit{I don't think seven emotions are sufficient. Emotion is complex and should be a spectrum with multiple dimensions. For example, the 'Junto Emotion Wheel' includes 64 distinct emotions.}" P01 said. P05 said that "\textit{There are many emotions, so it's unrealistic to represent them all, but I believe there should be an option to select two or three emotions to blend, helping to bridge gaps between different emotions.}"

We found that both non-asl hearing and DHH participants want to add more emotional voices to improve clarity in communication. \textbf{We believe, however, that expanding the range of emotions should be considered in conjunction with the capabilities of AI-generated emotional voice.} While users may wish to express a variety of emotions, it is essential for the system to reliably generate the corresponding emotional voice. We discuss this aspect further in the discussion section.

\subsubsection{Further Features for Emotional Voice}

To explore design implications for the translation system, an open-ended question was included about additional features for emotional voice.\textbf{ Hearing participants emphasized the importance of pitch, with 11 responses highlighting it as a key property for detecting emotional cues, along with volume and intensity.} Many noted that pitch serves as an effective indicator of specific emotions and is critical for identifying emotions. Concerns were raised about AI-generated voices sounding robotic or lacking natural inflection, with participants advocating for \textbf{more human-like tones and nuanced variations to enhance clarity}. The DHH group  emphasizes that incorporating more features into emotional voice systems will enhance the depth and subtlety of emotional expression. \textbf{DHH participants highlighted the importance of intensity to clarify the multiple meanings of words, as well as tone,} since ASL can vary in tone and intensity based on emotions. DHH03 noted,\textit{ "Intensity is important because words have multiple meanings, and intensity helps balance that."} DHH05 added, \textit{"Intensity and tone would be beneficial since Deaf people’s ASL can vary in tone and intensity based on their emotions, and this variation is present in every conversation."}

\section{Discussion}
Our study explores the integration of emotional voice cues into sign-to-speech translation systems as a novel means to enhance Deaf-to-Hearing communication. By addressing the emotional nuance often absent in existing translation tools, this work seeks to bridge a significant gap in collaborative interaction between DHH people and hearing individuals, fostering a more inclusive communication experience. Our findings underscore how emotional voice, as an augmentation to facial expressions in translation, supports clearer emotional comprehension, which is crucial for effective communication. Below, we discuss the implications for HCI and CSCW community, focusing on the collaborative experiences, design challenges, and opportunities for further development in multimodal translation tools.







\subsection{Challenges in Emotion Recognition through Sign Language}

Our results indicate that hearing participants often face cognitive and perceptual challenges when attempting to interpret signers’ emotions solely through facial expressions, especially when those expressions serve dual purposes. In American Sign Language (ASL), facial expressions do not simply convey emotions \cite{adolphs2002recognizing}; they also act as linguistic markers that modify the meaning of signs, such as expressing emphasis or intensity \cite{shaffer2018exploring, huenerfauth2011evaluating}. This multifunctionality can lead to confusion, as observed in our study, where participants frequently misinterpreted emotional states, especially when multiple facial expressions were used in a sequence. This finding underscores the complexity of sign language interpretation for non-signers and the potential benefits of incorporating complementary cues, such as emotional voice, to mitigate these challenges. Further studies could investigate ways to simplify emotion recognition by developing a system that explicitly indicates when a facial expression serves a linguistic purpose. This could involve adding a brief visual marker or an accompanying audio indicator. Additionally, exploring training modules for hearing users could help distinguish between emotional and linguistic expressions, potentially improving their interpretation accuracy and reducing cognitive confusion.

\subsection{Effectiveness and Limitations of Emotional Voice Integration}

The integration of emotional voice provided a notable improvement in participants' ability to accurately identify certain emotions. Specifically, emotions like anger, sadness, and fear were more accurately recognized with corresponding emotional voice cues than with a neutral voice or subtitles alone. This aligns with previous findings that suggest auditory cues are particularly effective in conveying strong or negative emotions, which are often more challenging to interpret visually \cite{nordstrom2019emotional, mccullough2005neural}. However, we observed that certain positive emotions, such as happiness, did not benefit as significantly from emotional voice cues. This could be due to subtleties in voice modulation that are less pronounced for positive emotions, suggesting the need for future systems to carefully calibrate positive emotional cues. To enhance the effectiveness of emotional voice cues, future research could involve fine-tuning voice modulation \cite{adigwe2018emotionalvoicesdatabasecontrolling} for positive emotions, potentially through a personalized adjustment feature that allows users to adjust voice tone based on their preferences. Additionally, conducting studies with a broader range of emotions and situational contexts could provide deeper insights into which emotions benefit most from voice cues and how voice modulation can be optimized for a wider emotional range.

\subsection{Cognitive Load and Information Processing in Multimodal Translation}

While emotional voice cues improved emotion recognition accuracy, they did not significantly reduce the cognitive load for participants \cite{creed2008psychological}. This may be attributed to the dual task of processing both sign content and emotional cues simultaneously. For hearing individuals unfamiliar with sign language, integrating visual and auditory information may require additional mental effort, as they are not only deciphering linguistic content but also discerning emotional intent. These findings highlight a potential trade-off in multimodal systems: while adding emotional voice aids comprehension, it also introduces a secondary processing task that could elevate cognitive load. Future systems could explore adaptive presentation of emotional cues, perhaps prioritizing emotional voice in contexts where the emotion is crucial to the message. To address cognitive load, future systems could experiment with adaptive algorithms that identify complex or ambiguous sign sequences and dynamically adjust the timing or prominence of emotional voice cues. Another promising avenue would be studying gradual user adaptation to multimodal cues and whether experience with the system can lead to improved efficiency in emotion recognition and reduced cognitive load over time.

\subsection{Design Implications for AI-based Sign-to-Speech Translation Systems}

Our study offers several implications for the design of future sign-to-speech systems. First, integrating emotional voice is promising, but systems should allow for adaptable emotional voice modulation to handle a broad spectrum of emotions, with careful attention to clarity, especially for positive emotions. Additionally, systems could explore alternative visual indicators that delineate linguistic markers from emotional expressions, potentially reducing the likelihood of misinterpretation by hearing users. Future systems could also incorporate real-time processing to recognize sequences of linguistic and emotional markers, creating smoother, contextually aware transitions between linguistic meaning and emotional tone. Such enhancements would better support users’ comprehension of a signer's intended message and emotional state. Future research could test various design elements, such as adaptive voice tone modulation and real-time emotional tone matching, in a range of conversational scenarios to refine system performance. Additionally, developing prototype systems with real-time feedback capabilities and assessing their impact on accuracy in emotion recognition could inform the development of user-centered translation systems that better address the diverse needs of DHH and hearing users.

\subsection{Generative AI Emotional Voices for Diverse Emotions}\label{emotionalvoice_discussion}

The number of AI-generated emotional voices that could be created depends on the sophistication of the AI model, the training data, and the specific parameters used to define emotions. At first, a model trained on a limited set of emotions—like joy, sadness, anger, fear, surprise, and disgust—might generate distinct voices for each emotion. This approach can be expanded to include more nuanced emotions like frustration, excitement, or contentment. Also, using dimensional models, such as Russell’s Circumplex Model \cite{russell1980circumplex}, an AI could create voices for emotions along axes like arousal (high to low energy) and valence (positive to negative). This could yield virtually limitless combinations, allowing for subtle gradations like “calm happiness” or “intense frustration.”. In addition, an AI can vary emotional voices by adjusting intensity and adding modifiers (e.g., “slightly sad” vs. “deeply sad”), which expands the range. Using these parameters, even a set of seven basic emotions could yield dozens of unique voices with varying intensities. Emotional expression can also adapt based on context (formal vs. informal) or the situation (urgent vs. casual). These adjustments add variety to each base emotion. With the combination of these factors, an advanced AI voice synthesis system could realistically generate hundreds or even thousands of emotional voices, depending on the granularity and context provided by the user. However, we should also consider how these emotional voices could be perceived by hearing users. Subtle emotional nuances might be interpreted differently, leading to potential confusion. Therefore, it may be necessary to evaluate the voices and prioritize those that score highly in clarity and accurately represent the corresponding emotions.

\subsection{Facial Expressions in ASL Beyond Emotions and Linguistic Markers}
We explored facial expressions limited to six emotions and six linguistic markers for study purposes. However, facial expressions in ASL extend beyond conveying emotions or linguistic markers in real-world contexts. They often serve pragmatic and discourse roles, like managing turn-taking, signaling emphasis, or showing hesitation. For instance, specific pauses or gaze shifts can imply a response is expected or stress a point. Additionally, ASL expressions can convey attitudes like friendliness, sarcasm, or urgency, adding context without altering signs’ lexical meaning. Future studies will explore non-signers’ perception of signers’ emotions with these varied expressions.
\section{Conclusion}

For a long time, American Sign Language (ASL) translation systems have only produced neutral voices when converting signs to speech for hearing individuals. However, due to the complex nature of ASL, which incorporates both emotional facial expressions and linguistic markers, hearing individuals may misinterpret the intended emotions of the signer, even if the translated message is accurate. Our research indicates that linguistic markers are often misinterpreted as negative emotions, potentially hindering effective communication between signers and hearing individuals.

To address this issue, we proposed an emotional ASL translation system that integrates emotional voice to reduce misunderstandings of both emotional facial expressions and linguistic markers. Our findings demonstrate that when emotional voice is provided, the accuracy of recognizing both emotions and conveyed messages improves, compared to using no voice or neutral voice assistance.

Furthermore, our research offers valuable insights for the design of emotional voice systems. The results show that our system is particularly effective in conveying emotions such as anger, sadness, and surprise. Feedback from both hearing and DHH participants highlights important voice characteristics, including pitch variation, intensity, volume, and a more personalized tone, which could further enhance the effectiveness of emotional voice systems.

\bibliographystyle{ACM-Reference-Format}
\bibliography{main}

\end{document}